\documentclass[epj]{webofc}
\usepackage[utf8]{inputenc}
\usepackage[varg]{txfonts}   
\usepackage{booktabs}
\usepackage{xcolor}
\definecolor{darkred}{rgb}{0.4,0.0,0.0}
\definecolor{darkgreen}{rgb}{0.0,0.4,0.0}
\definecolor{darkblue}{rgb}{0.0,0.0,0.4}
\usepackage[bookmarks,linktocpage,colorlinks,
    linkcolor = darkred,
    urlcolor  = darkblue,
    citecolor = darkgreen]{hyperref}
\usepackage{subfigure}
\usepackage{bm}
\usepackage{mathrsfs}
\usepackage{dcolumn}
\wocname{EPJ Web of Conferences}
\woctitle{Lattice2017}
\newcommand\Dl{\raisebox{0.05em}{$\stackrel{\scriptstyle\leftarrow}D$}}
\newcommand{\subrm}[1]{{\scriptscriptstyle\mathrm{#1}}}
\newcolumntype{d}[1]{D{.}{.}{#1} }
\begin{document}
%
\selectlanguage{english}
\title{%
Improving the theoretical prediction for the $\bm{B_s-\bar{B}_s}$ width difference: 
matrix elements of next-to-leading order $\bm{\Delta B=2}$ operators 
}
\author{%
\firstname{Christine} \lastname{Davies}\inst{1}\fnsep \and
\firstname{Judd} \lastname{Harrison}\inst{2} \and
\firstname{G Peter} \lastname{Lepage}\inst{3} \and
\firstname{Christopher} \lastname{Monahan}\inst{4,5} \and
\firstname{Junko} \lastname{Shigemitsu}\inst{6} \and
\firstname{Matthew}  \lastname{Wingate}\inst{2}\fnsep\thanks{Speaker, \email{M.Wingate@damtp.cam.ac.uk}}
}
\institute{%
  SUPA, School of Physics and Astronomy, University of Glasgow, Glasgow
  G12 8QQ, United Kingdom
\and
DAMTP, University of Cambridge, Cambridge CB3 0WA, United Kingdom
\and
Laboratory of Elementary Particle Physics, Cornell University, Ithaca, NY
14853, United States
\and
New High Energy Theory Center and Department of Physics and Astronomy,
Rutgers, the State University of New Jersey, Piscataway, NJ 08854, United
States
\and
Institute for Nuclear Theory, University of Washington, Seattle, WA 98195-1550,
United States
\and
Department of Physics, Ohio State University, Columbus, Ohio 43210,
United States
}
\abstract{
  We present lattice QCD results for the matrix elements of
  $R_2$ and other dimension-7, $\Delta B = 2$ operators relevant for
  calculations of $\Delta \Gamma_s$, the $B_s-\bar{B}_s$ width
  difference.  We have computed correlation functions using 5
  ensembles of the MILC Collaboration's 2+1+1-flavour gauge field
  configurations, spanning 3 lattice spacings and light sea quarks
  masses down to the physical point.  The HISQ action is used for the
  valence strange quarks, and the NRQCD action is used for the bottom
  quarks.  Once our analysis is complete, the theoretical uncertainty
  in the Standard Model prediction for $\Delta \Gamma_s$ will be
  substantially reduced.
}
\maketitle
\section{Introduction}
\label{sec:intro}

Mixing between particle and antiparticle states of neutral mesons has
now been observed in $K^0$, $D^0$, $B^0$, and $B_s^0$ mesons.  These
mixings are due to couplings between generations of quark
$\mathrm{SU}(2)_L$ doublets after electroweak symmetry breaking.
Since the leading-order weak process, represented by the ``box''
diagrams, is at 1-loop level, there is the chance that new heavy
particles, beyond those in the Standard Model, could cause differences
between Standard Model predictions and experimental results.

To a good approximation, three parameters are sufficient to describe
neutral meson mixing: the moduli of the off-diagonal matrix elements
of the $2\times 2$ mass and width matrices, $M$ and $\Gamma$, and
their relative phase $\phi = \arg(-M_{12}/\Gamma_{12})$.  For the
$B_s^0$ system these parameters are constrained by experimental
measurements of the $B_s^0$-$\bar{B}_s^0$ mass difference, width
difference, and a flavour-specific CP asymmetry:
\begin{equation}
  \Delta M_s = 2|M_{12}^s|\,, ~~ \Delta\Gamma_s = 2|\Gamma_{12}^s|\cos\phi_s \,,
  ~\mbox{and}~ a_{\mathrm{fs}}^s
  = \frac{\Delta \Gamma_s}{\Delta M_s}
  \tan\phi_s \,.
  \label{eq:mixing_eqns}
\end{equation}
In (\ref{eq:mixing_eqns}) and the remainder of this paper we use
notation specific to $B_s^0$ mixing.  See Ref.~\cite{Artuso:2015swg}
for a recent review of $B_s^0$ mixing and references to a rich
literature.

The calculation of $\Delta\Gamma_s$ within the Standard Model
is summarized in Ref.~\cite{Lenz:2006hd}.  
Contributions to $\Gamma_{12}^s$ come from matrix elements of the
non-local product of 2 $\Delta B=1$ effective Hamiltonians
\begin{equation}
  \mathscr{T} = \mathrm{Im} \,i\!\int\!d^4x\, \mathcal{T} \,
  H_{\mathrm{eff}}^{\Delta B=1}(x) \, H_{\mathrm{eff}}^{\Delta B=1}(0)\,.
  \label{eq:nonlocal}
\end{equation}
The contributions from charm and up quarks in the intermediate
states depend on the corresponding CKM matrix elements; i.e.\
$\Gamma_{12}^s = -(\lambda_c^2 \Gamma_{12}^{cc} + 2 \lambda_c\lambda_u
\Gamma_{12}^{uc} + \lambda_u^2\Gamma_{12}^{uu})$ with $\lambda_i =
V_{is}^* V_{ib}$.  At the present level of accuracy, only the CKM-leading
contribution from $\Gamma_{12}^{cc}$ is important.

Direct lattice QCD calculation of matrix elements such as $\langle
\bar{B}_s| \mathscr{T} | B_s \rangle$ is not generally feasible due to
the difficulty in correctly treating all intermediate
states.\footnote{Progress is being made in the kaon system with
  heavier than physical quark masses, where the only intermediate
  state with energy less than $m_K$ is the $\pi^0$
  \cite{Bai:2014cva}.}  However, one can employ an operator product
expansion known as the heavy quark expansion (HQE).  Order-by-order in
$\Lambda_{QCD}/m_B$, one relates the matrix elements of nonlocal
operators to a series of matrix elements of local $\Delta B =2$
operators. Using the most advantageous basis \cite{Lenz:2006hd}
the charm-charm loop contribution to $\Gamma_{12}^s$ is given by
\begin{equation}
  \Gamma_{12}^{cc} = \frac{G_F^2 m_b^2}{24 \pi m_{B_s}}
  \left[\left(G + \frac{\alpha_2}{2} G_S\right)
    \langle \bar{B}_s | Q_1 | {B}_s \rangle 
    + \alpha_1 G_S \langle \bar{B}_s | Q_3 | {B}_s \rangle\right] \;+\; 
  \tilde{\Gamma}_{12,1/m_b}^{cc} \,,
\end{equation}
with the next order in the HQE given by
\begin{equation}
  \tilde{\Gamma}_{12,1/m_b}^{cc} = \frac{G_F^2 m_b^2}{24 \pi m_{B_s}}
  \left\{ g_0^{cc}\langle \bar{B}_s | R_0 | {B}_s \rangle
  + \sum_{j=1}^3\left[g_j^{cc} \langle \bar{B}_s | R_j | {B}_s \rangle
    + \tilde{g}_j^{cc} \langle \bar{B}_s | \tilde{R}_j | {B}_s \rangle\right]  
  \right\} \,.
\end{equation}

A full basis of dimension-6 $\Delta B=2$ operators
can be written as
\begin{alignat}{4}
  Q_1 & = (\bar{b}^\alpha \gamma^\mu(1-\gamma^5)s^\alpha)(\bar{b}^\beta
  \gamma_\mu(1-\gamma^5) s^\beta)\,,  & ~~~~
  Q_4 & = (\bar{b}^\alpha (1-\gamma^5) s^\alpha)(\bar{b}^\beta
  (1+\gamma^5) s^\beta) \nonumber \\
  Q_2 &= (\bar{b}^\alpha (1-\gamma^5) s^\alpha)(\bar{b}^\beta (1-\gamma^5) s^\beta)
  \,, & ~~~~ Q_5 &= (\bar{b}^\alpha (1-\gamma^5) s^\beta)(\bar{b}^\beta
  (1+\gamma^5) s^\alpha) \nonumber \\
  Q_3 &= (\bar{b}^\alpha (1-\gamma^5) s^\beta)(\bar{b}^\beta (1-\gamma^5)s^\alpha)
  \,.
  \label{eq:Qops}
\end{alignat}
At higher order in the HQE, one needs matrix elements of the following
operators
\begin{align}
  R_0 &= Q_2 + \alpha_1 Q_3 + \frac12 \alpha_2 Q_1 \nonumber \\
  R_1 & = \frac{m_s}{m_b}(\bar{b}^\alpha(1-\gamma^5)s^\alpha)
  (\bar{b}^\beta (1+\gamma^5)s^\beta) = \frac{m_s}{m_b} Q_4 \nonumber \\
  R_2 & = \frac{1}{m_b^2} (\bar{b}^\alpha \Dl_\rho \gamma^\mu(1-\gamma^5)
  D^\rho s^\alpha)
  (\bar{b}^\beta \gamma_\mu(1-\gamma^5)s^\beta) \nonumber \\
  R_3 & = \frac{1}{m_b^2} (\bar{b}^\alpha \Dl_\rho (1-\gamma^5) D^\rho s^\alpha)
  (\bar{b}^\beta (1-\gamma^5)s^\beta) \,.
  \label{eq:Rops}
\end{align}
Matrix elements of the $Q_i$ operators (\ref{eq:Qops}) have long been
calculated using lattice QCD; unquenched results for $B_s$ mixing
appear in
\cite{Gamiz:2009ku,Carrasco:2013zta,Dowdall:2014qka,Bazavov:2016nty}.
Until this work there have been no calculations of $R_2$ and $R_3$
matrix elements.  In phenomenological analyses
\cite{Lenz:2006hd,Jubb:2016mvq} the vacuum saturation approximation
was used, allowing a 50\% uncertainty.  Sum rule estimates suggest
these matrix elements should be within a few percent of the vacuum
saturation approximation values \cite{Mannel:2007am}, although the VSA
predictions depend sensitively on the value of the $b$-quark pole
mass.

\section{Method}
\label{sec:method}

\begin{table}[thb]
  \small
  \centering
  \caption{Parameters of the ensembles used in this calculation.}
  \label{tab:params}
  \begin{tabular}{cd{12}d{5}d{4}d{3}cd{4}d{3}}\toprule
    label & \multicolumn{1}{c}{$a$/fm}
    & \multicolumn{1}{c}{$am_\ell^{\mathrm{sea}}$} &
    \multicolumn{1}{c}{$am_s^{\mathrm{sea}}$} &
    \multicolumn{1}{c}{$am_c^{\mathrm{sea}}$} &
    $N_s^3 \times N_t$ & \multicolumn{1}{c}{$am_s^{\mathrm{val}}$} &
    \multicolumn{1}{c}{$am_b$}
    \\\midrule
    VC5 & 0.1474(5)(14)(2) & 0.013 & 0.0650 & 0.838 & $16^3\times 48$   &
    0.0641 & 3.297 \\
    VCp & 0.1450(3)(14)(2) & 0.00235 & 0.0647 & 0.831 & $32^3\times 48$ &
    0.0628 & 3.25  \\
    C5 & 0.1219(2)(9)(2)  & 0.0102 & 0.0509 & 0.635 & $24^3\times 64$  &
    0.0522 & 2.66  \\
    Cp & 0.1189(2)(9)(2)  & 0.00184 & 0.0507 & 0.628 & $48^3\times 64$ &
    0.0507 & 2.62 \\
    F5 & 0.0873(2)(5)(1)  & 0.0074 & 0.037 & 0.440 & $32^3\times 96$   &
    0.0364 & 1.91
  \\\bottomrule
  \end{tabular}
\end{table}

We use MILC's HISQ ensembles, which include sea quark effects of degenerate
up and down quarks and physical-mass strange and charm quarks  
\cite{Bazavov:2010ru,Bazavov:2012xda}.  We use
the HISQ action for the valence $s$ quark and the NRQCD action
for the $b$ quark.  The 5 ensembles include 3 distinct lattice spacings 
which we respectively refer to as fine (F), coarse (C), and 
very coarse (VC).  For each of these spacings we use configurations with
dynamical pion mass of approximately 300 MeV, and for the coarse and
very coarse spacings, we used the physical ensembles which have pion
mass approximately 130 MeV.  Table~\ref{tab:params} lists specific input
parameters and the lattice spacings as determined from the $\Upsilon(2S-1S)$
splitting \cite{Dowdall:2011wh,Dowdall:2013tga}.

In carrying out the calculation of $\langle\bar{B}_s|R_i|B_s\rangle$,
with $i=2,3$, we need not compute all 4 terms in the Lorentz dot
product (\ref{eq:Rops}).  The temporal derivative acting on the $b$
field is $O(m_b)$: $\bar{b} \Dl_0 = \pm m_b \bar{b} \gamma_0$, the
sign depending on whether we have an outgoing $b$ quark or incoming
$\bar{b}$ antiquark.  Thus we can write
\begin{equation}
  \frac{1}{m_b^2}(\bar{b}^\alpha \Dl_\rho \Gamma D^\rho s^\alpha)
  = \frac{1}{m_b^2}(\bar{b}^\alpha \Dl_0 \Gamma D^0 s^\alpha)
  + O\left(\frac{1}{m_b^2}\right) \,.
\end{equation}
Applying the equations of motion, $i\gamma_0 D^0 s =
(\vec{\gamma}\cdot \vec{D})s$, we arrive at
\begin{equation}
R_{2,3} = \pm\frac{1}{m_b}(\bar{b}^\alpha \Gamma \gamma_0 
(\vec{\gamma}\cdot \vec{D})s^\alpha)(\bar{b}^\beta \Gamma s^\beta) \,.
\end{equation}

The lattice calculation of $\langle\bar{B}_s|R_{2,3}|B_s\rangle$
proceeds just as for the $Q_j$ matrix elements, except for the need to have
a derivative operate on the strange quark at the operator.  
In addition to needing a staggered propagator $g(y,z)$ computed from local
source \cite{Wingate:2002fh}
\begin{equation}
K(x,y) \, g(y,z) = \delta(x,z) \,,
\end{equation}
we need propagators from a point-split source ($k=1,2,3$)
\begin{equation}
K(x,y) \,g^{(k)}(y,z) = \frac12\left[\delta(x,z+\hat{k}a) U_k^\dagger(z)
- \delta(x, z-\hat{k}a) U_k(z-\hat{k}a)\right] \,.
\end{equation}
Naive quark propagators are constructed from staggered propagators via
\begin{align}
G(y,z) & = \Omega(y) \, g(y,z) \, \Omega^\dagger(z) \nonumber \\
G^{(k)}(y,z) & = \Omega(y) \, g^{(k)}(y,z) \,\Omega^\dagger(z\pm \hat{k}a) 
\end{align}
where $\Omega(x) = \prod_{\mu=0}^3 (\gamma_\mu)^{x_\mu/a}$.
Since we will need to sum over spatial directions, we require 4 strange quark
inversions on each configuration and for each source location.

\section{Perturbative matching}
\label{sec:matching}

One-loop matching between the lattice theory and the continuum
$\overline{\mathrm{MS}}$ renormalization schemes has been carried out for the
$Q_i$ operators, including tree-level $1/m_b$ corrections
\cite{Monahan:2014xra}.  The 1-loop mixing between operators is
parametrized by the $\rho_{ij}$ matrix and $1/m_b$ corrections are
given by $\hat{Q}_{i,1}^{\mathrm{sub}}$, which are of the form
$\frac{1}{2m_b}(D_k \bar{b}^\alpha \gamma^k \Gamma_1
s^\alpha)(\bar{b}^\beta \Gamma_2 s^\beta)$:
\begin{equation}
  Q_i^{\overline{\subrm{MS}}} = \hat{Q}_i + \alpha_s \rho_{ij}
  \hat{Q}_j + \hat{Q}_{i,1}^{\mathrm{sub}} \,.
\label{eq:match_msbar}
\end{equation}
Because derivatives are implemented as finite difference operators
the $\frac{1}{a}\hat{Q}_{i}$ mix with $\hat{Q}_{i,1}$; this can similarly
be computed in perturbation theory.  We define a subtracted operator
which gives a more accurate determination of the next-to-leading
contribution:
\begin{equation}
   \hat{Q}_{i,1}^{\mathrm{sub}} = \hat{Q}_{i,1} - \alpha_s \zeta_{ij} \hat{Q}_j \,.
\label{eq:match_subtraction}
\end{equation}
The coefficients $\rho_{ij}$ and $\zeta_{ij}$ are tabulated in
\cite{Monahan:2014xra}.

A similar subtraction is done here for the $R_{2,3}$ operators:
\begin{equation}
  \hat{R}_i^{\mathrm{sub}} = \hat{R}_i - \alpha_s \xi_{ij} \hat{Q}_j \,.
 \label{eq:match_Rsub}
\end{equation}
Values for $\xi_{ij}$ are given in Table~\ref{tab:perturb}.  We use
the $\alpha_V$ values as tabulated in \cite{Colquhoun:2015oha}.  Note that we
have not carried out the 1-loop matching between lattice and
$\overline{\mathrm{MS}}$ schemes for $\hat{Q}_{i,1}^{\mathrm{sub}}$
or $\hat{R}_i^{\mathrm{sub}}$.  Therefore our results for their
$\overline{\mathrm{MS}}$ matrix elements will have an $O(\alpha_s)$
systematic uncertainty.

\begin{table}
  \small
  \centering
  \caption{\label{tab:perturb}  Perturbative subtraction coefficients
    used in (\ref{eq:match_Rsub}), for the values of $am_b$ used on
    each ensemble.
    }
  \begin{tabular}{cd{5}d{5}d{5}d{5}d{5}} \toprule
    Coefficient & \multicolumn{1}{c}{VC5} & \multicolumn{1}{c}{VCp} &
    \multicolumn{1}{c}{C5} & \multicolumn{1}{c}{Cp} & \multicolumn{1}{c}{F5}
    \\ \midrule
    $\xi_{21}$ & -0.1311 & -0.1327 & -0.1557 & -0.1573 & -0.2004 \\ 
    $\xi_{22}$ & 0.0092 & 0.0093 & 0.013 & 0.0133 & 0.0225 \\ 
    $\xi_{31}$ & -0.0331 & -0.0334 & -0.0392 & -0.0397 & -0.0508 \\ 
    $\xi_{32}$ & -0.2829 & -0.2864 & -0.3404 & -0.3449 & -0.451 \\
    \bottomrule
  \end{tabular}
\end{table}

\section{Fits to correlation functions}
\label{sec:corr}

On each of the 1000 or so configurations in the 5 ensembles listed in
Table~\ref{tab:params}, we created strange quark propagators with
inversion sources on 2 timeslices per configuration -- except for the
VC5 ensemble where we weighed the benefits of doubling the number of
sources per configuration.\footnote{We concluded that increased
  statistics were not sufficiently beneficial to warrant the cost of
  doubling the data set on other ensembles.}  We calculated 3-point
functions with local $B_s$ and $\bar{B}_s$ sinks as well as Gaussian
smeared sinks with 2 radii.  The smearing was done with the links
fixed to Coulomb gauge.  The 2-point correlators are taken from
earlier work where 16 sources per configuration were used
\cite{Dowdall:2013tga}.

Correlator data are fit to functions of the form
\begin{align}
  C^{\mathrm{2pt}}_{ab}(t) & = \sum_i X_{a,i} X_{b,i} e^{-E_i t}
  + \sum_i (-1)^{t/a} Y_{a,i} Y_{b,i} e^{-E_i^o t}\\
  \label{eq:C2pt}
  \intertext{and}
  C^{\mathrm{3pt}}_{ab}(t,T) & = \sum_{i,j} X_{a,i} V_{nn,ij} X_{b,j}
  e^{-E_i t} e^{-E_j(T-t)} ~+~\mathrm{oscillating}
  \label{eq:C3pt}
\end{align}
using the \texttt{corrfitter} package \cite{Lepage:corrfitter}.
The oscillating states in (\ref{eq:C2pt}) and (\ref{eq:C3pt}) appear due to
opposite-parity temporal doublers present in staggered quark formulations.
In (\ref{eq:C3pt}), $t$ is the temporal distance between the initial
state interpolating operator and the 4-quark operator and $T$ is the
distance between the initial and final state interpolating operators.
Values used in the fits presented here are given in Table~\ref{tab:tranges}.

\begin{table}
  \small
  \centering
  \caption{\label{tab:tranges} Ranges in Euclidean time used for
    fits to correlation functions.  Numbers are given in lattice
    units.}
  \begin{tabular}{cccc} \toprule
    Ensemble(s) & $t_{\mathrm{min}}$ & $t_{\mathrm{max}}^{\mathrm{2pt}}$ & $T$
    \\ \midrule
     VC5, VCp & 5 & 17 & 11, 12, 13  \\ 
     C5, Cp & 6 & 21 & 13, 14, 15, 16 \\ 
     F5 & 9 & 40 & 19, 20, 21, 22, 23, 24, 25 \\
    \bottomrule
  \end{tabular}
\end{table}

\begin{figure}[t]
  \centering
  \includegraphics[width=0.47\textwidth]{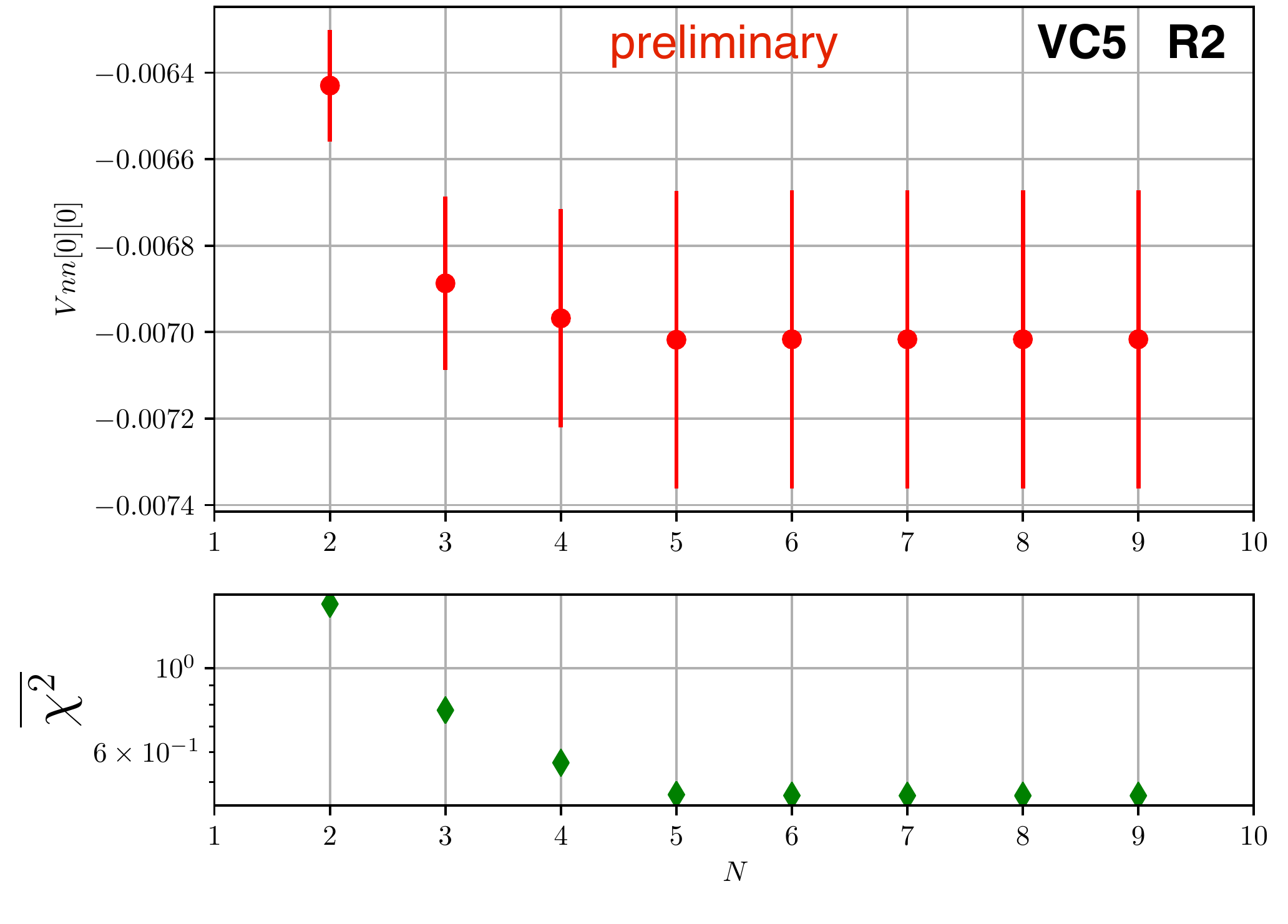}
  \includegraphics[width=0.47\textwidth]{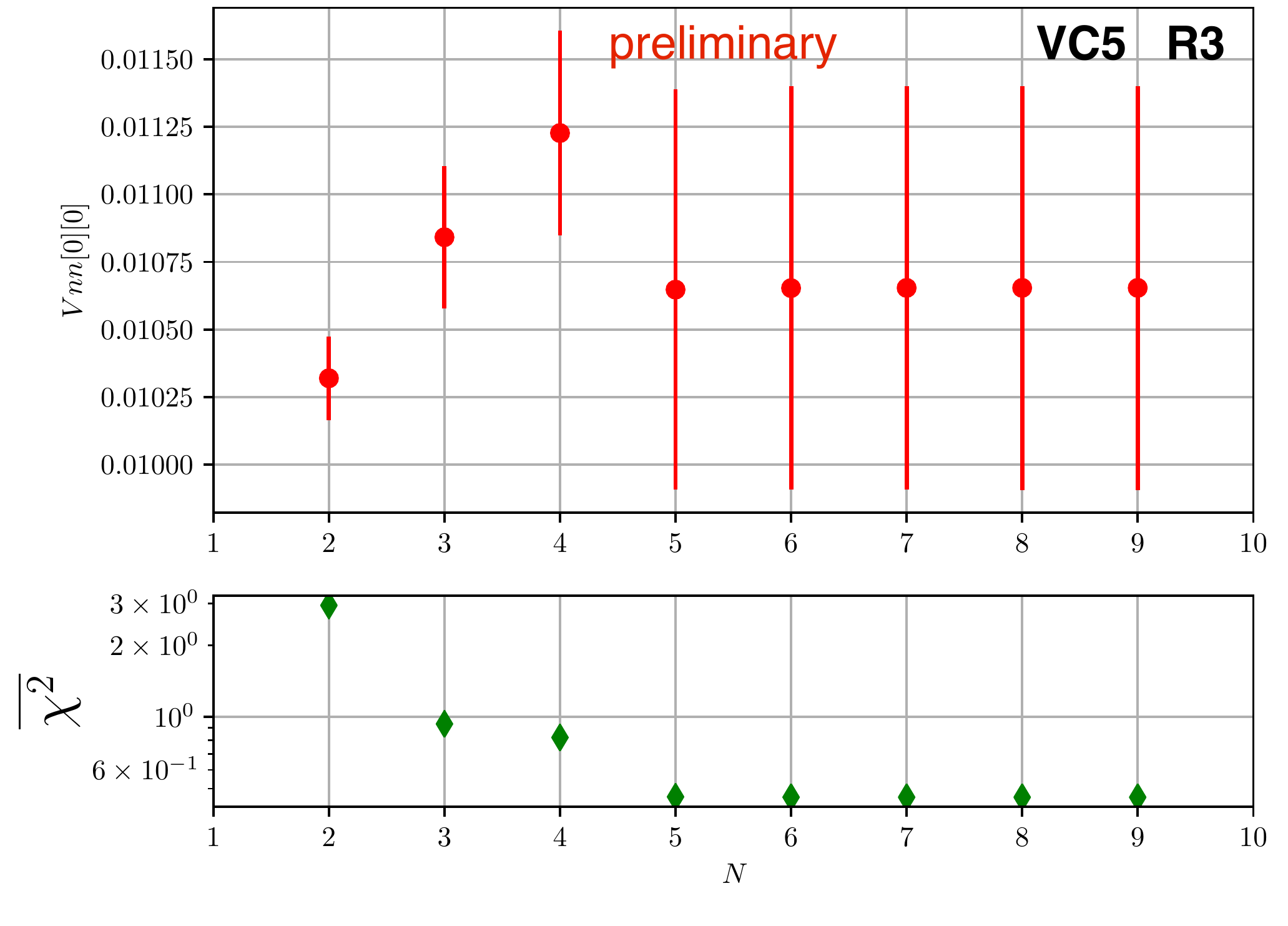}
  \includegraphics[width=0.47\textwidth]{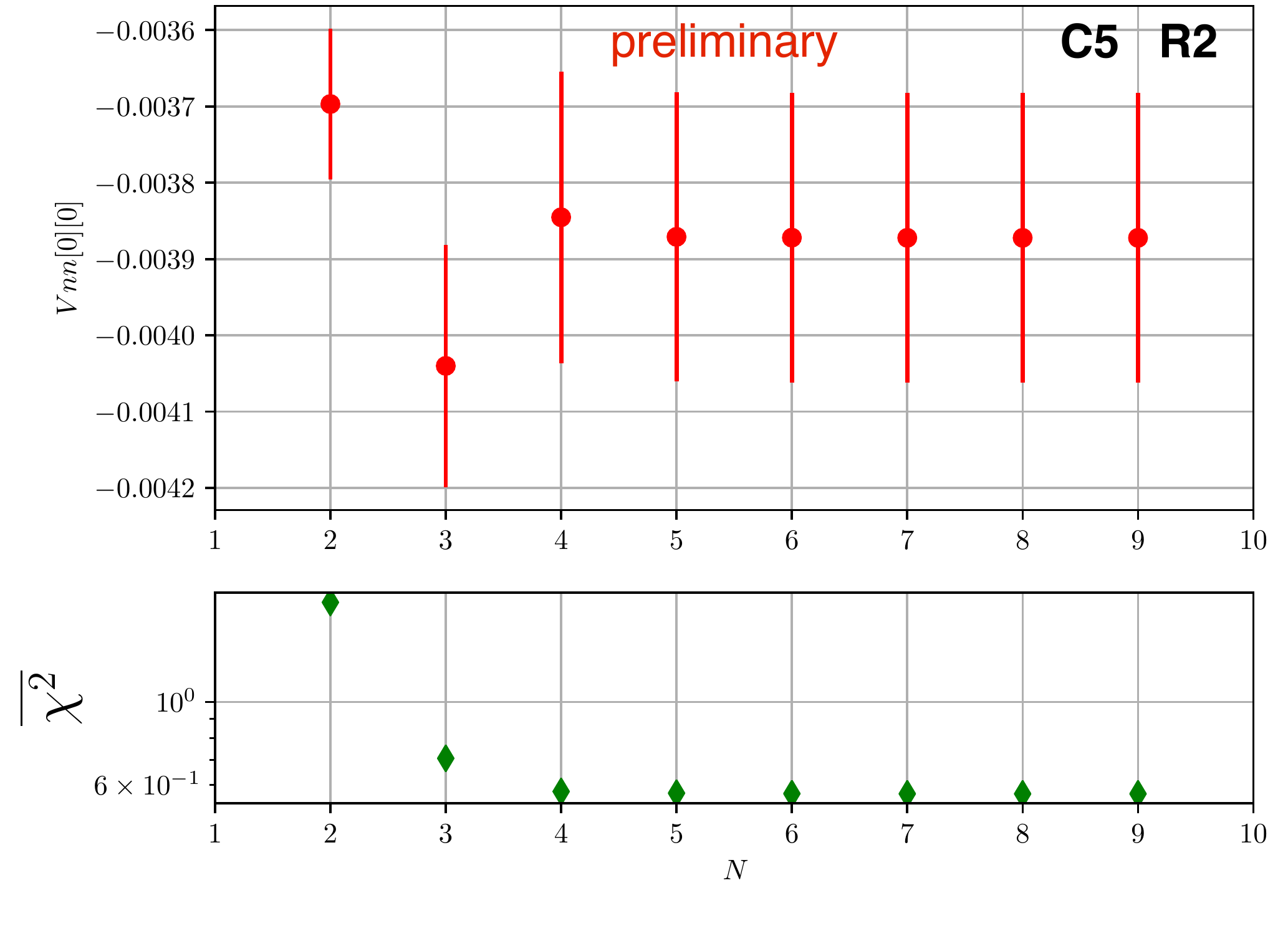}
  \includegraphics[width=0.47\textwidth]{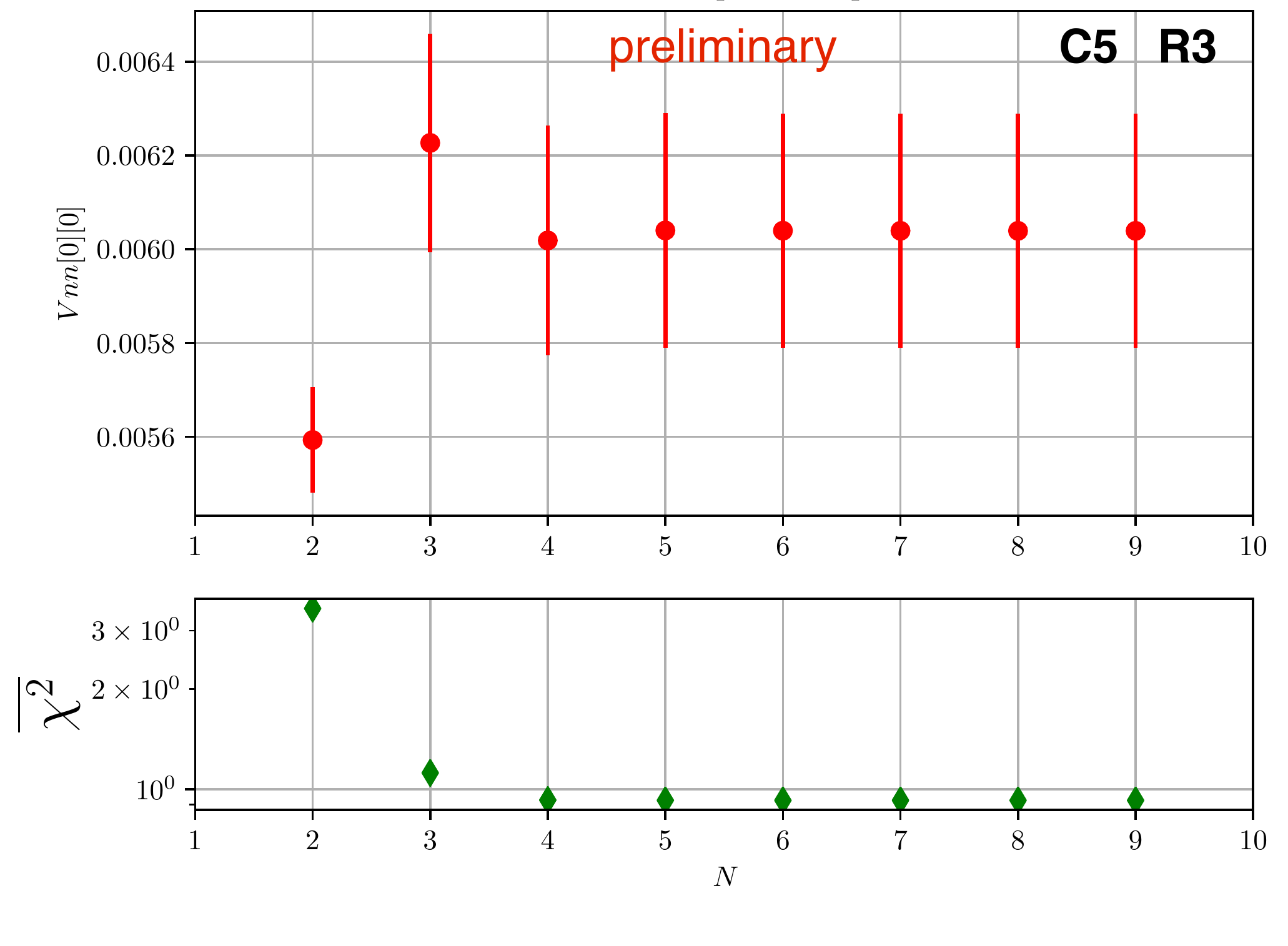}
  \includegraphics[width=0.47\textwidth]{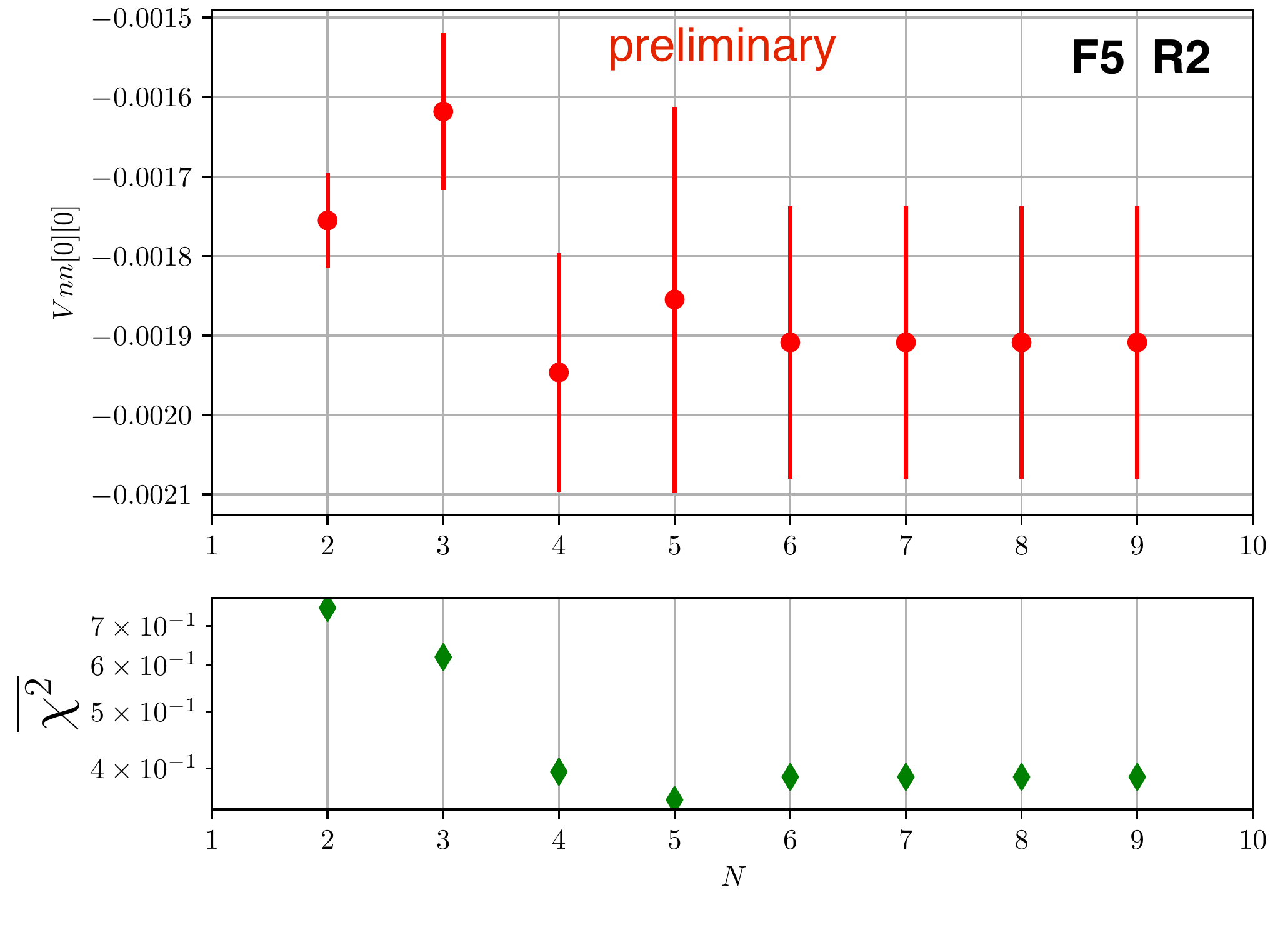}
  \includegraphics[width=0.47\textwidth]{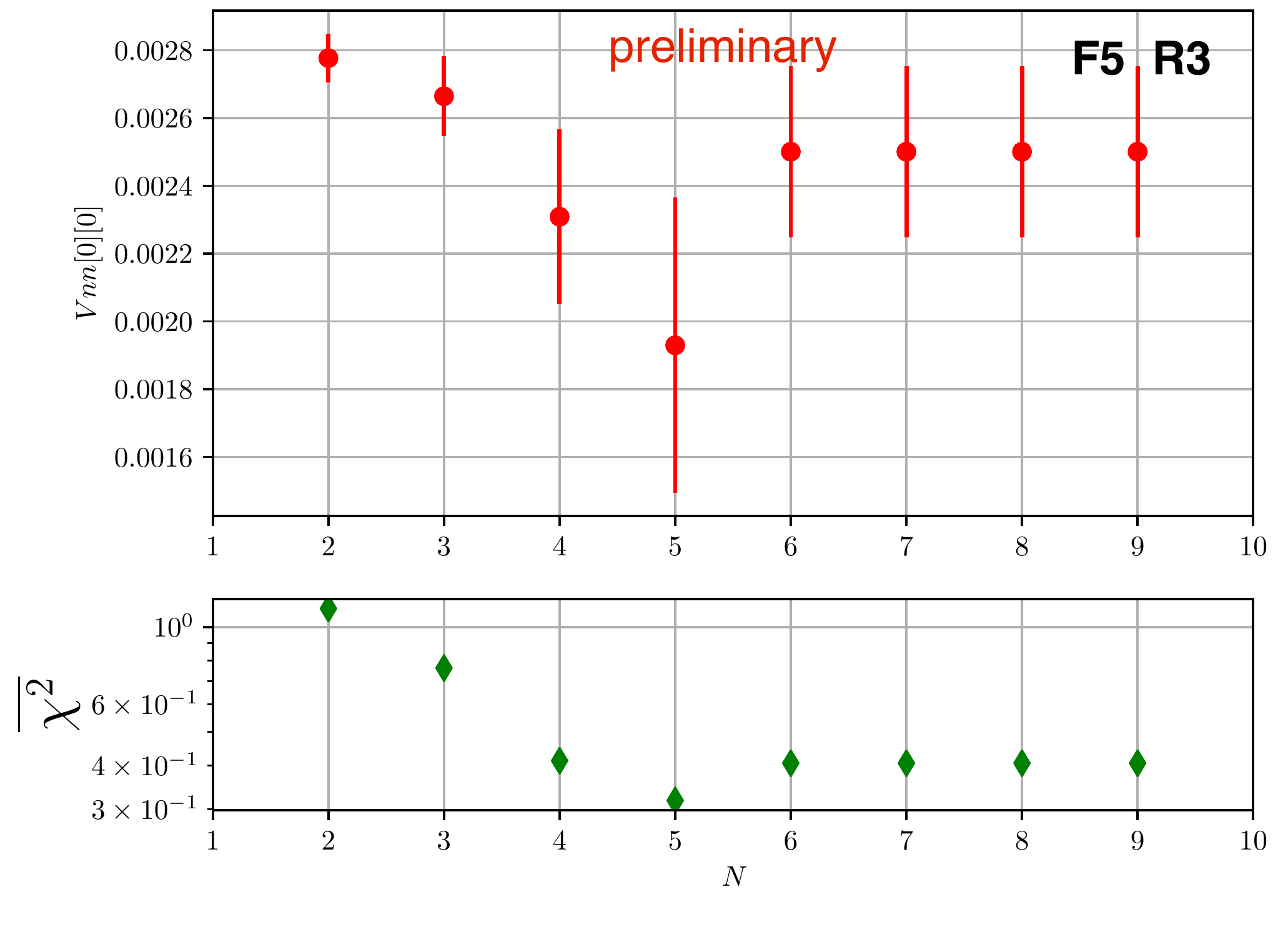}
  \caption{Fit results (upper plots) and $\chi^2$ per
    degree-of-freedom (lower plots) for unsubtracted $R_2$ and $R_3$
    fit amplitudes ($V_{nn,00}$) for increasing number of exponentials
    (Eq.~\ref{eq:C3pt}) on the VC5, C5, and F5 ensembles. $N$ is equal
    to the number of energies in the nonoscillating channel (desired parity)
    and the number of energies in the oscillating channel.}
  \label{fig:fitsNexp}
\end{figure}

The Gaussian priors for the fit amplitudes and energies were set as
follows.  We first performed 2-exponential fits ($N=2$ exponentials in
each parity channel) to the 2-point data using wide priors and
$t_{\mathrm{min}} \gtrsim 1.2$ fm.  From the output of this fit we took
the ground state energy and amplitude, multiplied their uncertainties
by 10, and used this as the prior means and half-widths for subsequent
fits.  For the excited states, we took the energy splittings to be
$O(a\Lambda_{\subrm{QCD}}) \pm 50\%$ and the amplitudes to be $0\pm 1$.

After fixing the priors for the energies and 2-point amplitudes, we
performed $N=3$ fits to 3-point correlator data with $t_{\mathrm{min}}\approx
1.0$ fm and 2 large values of $T$ using $0\pm 1$ for the priors on the
$V$ fit parameters.  This gave an order-of-magnitude estimate for the
ground state amplitude.  In subsequent fits we set the prior on $V_{nn,00}$
to be the fit result $\pm 50-100\%$; for the amplitudes in the oscillating
terms, we used standard deviations of $100-400\%$ of the results from
the preliminary fits.

In the fits presented below, we found that convergence was improved by
first fitting the 2-point correlator data and using the results as
priors for the fits to the 3-point correlators.  In most cases the
difference between these ``chained'' fits and fully simultaneous fits
is not significant \cite{Bouchard:2014ypa}; however, there were some
cases where the simultaneous fits failed to converge.

In Fig.~\ref{fig:fitsNexp} we show preliminary results of fits to the
3-point amplitudes $V_{nn,00}$ determining the $R_2$ and $R_3$ matrix
elements.  We observe results which give consistent results once
enough exponentials are included to account for excited state contributions
to the correlators.  In order to obtain this, it was necessary to
impose an SVD cut of 0.001 on the correlation matrix whose smaller
singular values are not well-determined by the data.

\begin{figure}
  \centering
  \includegraphics[width=0.49\textwidth]{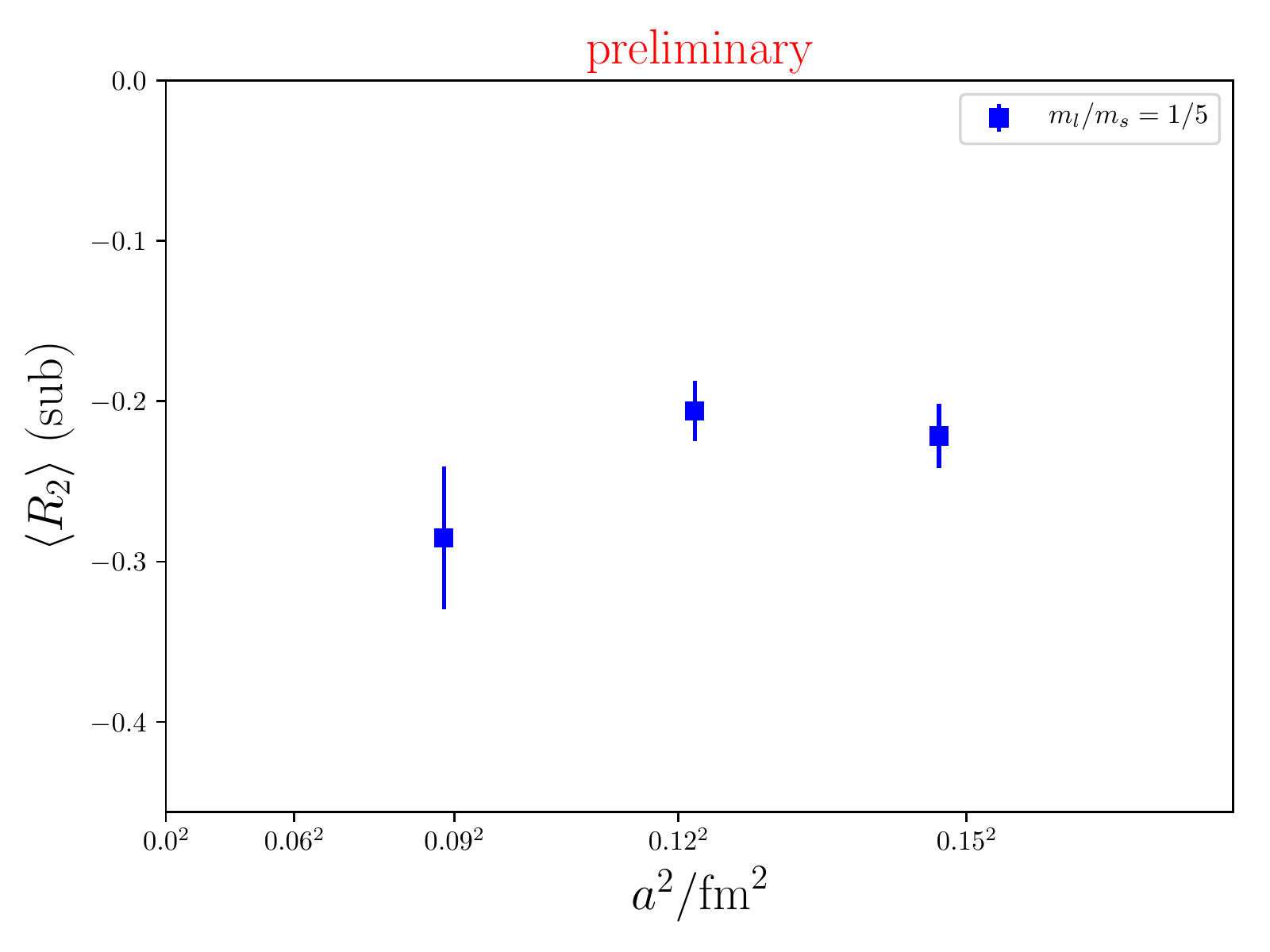}
  \includegraphics[width=0.49\textwidth]{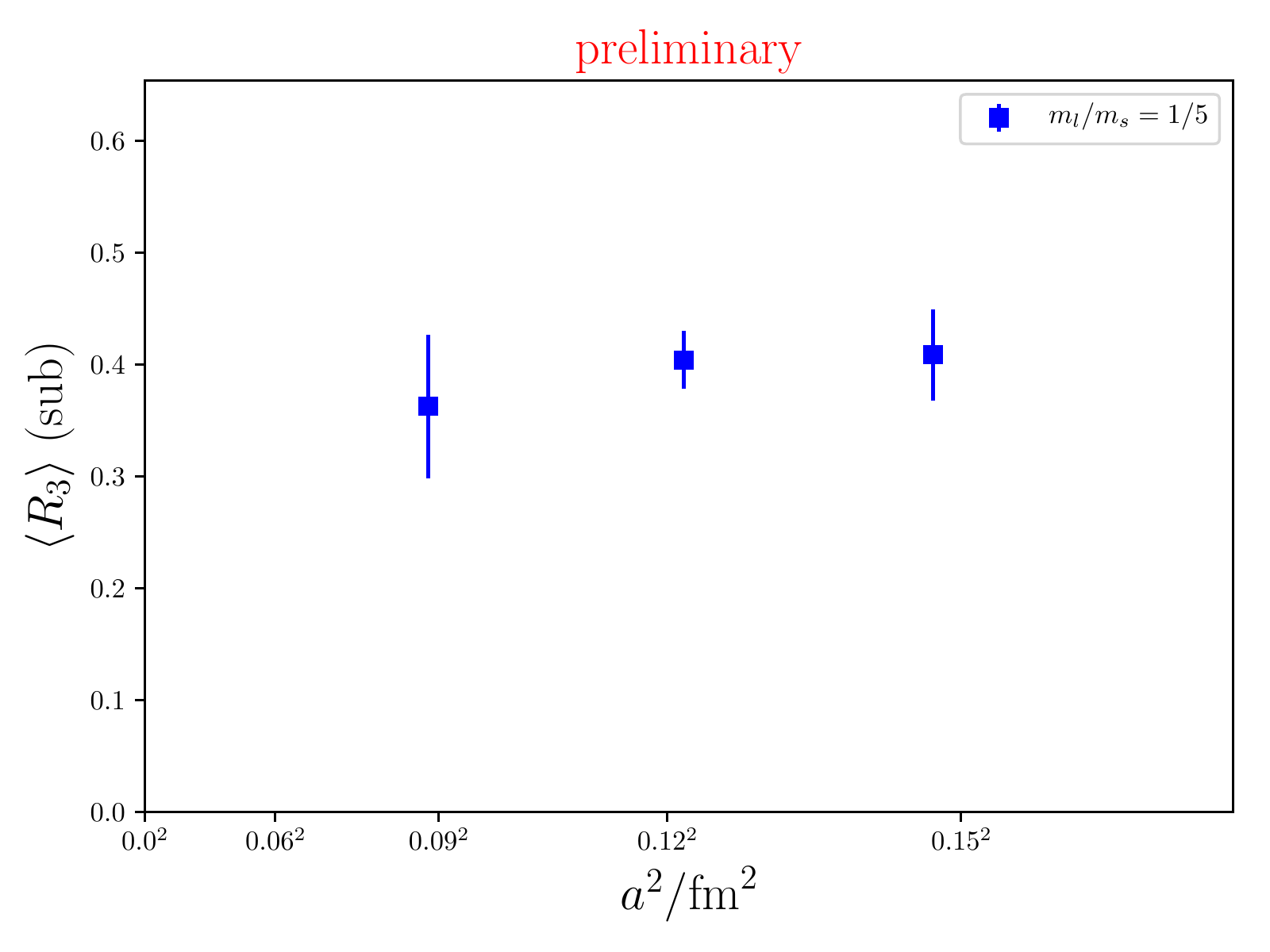}
  \caption{Results (in GeV${}^4$) for subtracted $R_2$ and $R_3$
    matrix elements on the ensembles with $m_\ell/m_s = 1/5$.  Error
    bars shown only include statistical and fitting uncertainties.
    The vacuum saturation approximation gives $\langle R_2\rangle^{\subrm{VSA}} =
    -0.3$ GeV${}^4$ and $\langle R_3\rangle^{\subrm{VSA}}=0.5$ GeV${}^4$.}
  \label{fig:r_vs_asq}
\end{figure}

Figure~\ref{fig:r_vs_asq} shows preliminary results vs.\ $a^2$ for matrix
elements of $R_2$ and $R_3$ after subtraction (\ref{eq:match_Rsub}),
on the VC5, C5, and F5 ensembles.  The fits on the physical point
ensembles VCp and Cp are not as far along in the process of being
checked.  We are still assessing fitting uncertainties and ensuring
results are robust against different fitting choices.  What we present
here are the results of separate fits to correlators for the operators
$R_2$, $R_3$, $Q_1$, and $Q_2$.  Once we have finished investigating
these fits, we will form the linear combinations of correlators,
configuration-by-configuration, which will allow us to determine
matrix elements of $R_2^{\mathrm{sub}}$ and $R_3^{\mathrm{sub}}$
directly.

\section{Outlook}

We presented preliminary results for $\langle \bar{B}_s |R_2|B_s\rangle$
and $\langle \bar{B}_s |R_3|B_s\rangle$ on 5 ensembles spanning a range
of 3 lattice spacing and including 2 physical mass ensembles.  We are
presently verifying stability of fit results.  The statistical precision
may be improved by performing fits to the linear combinations of correlators
directly yielding the 1-loop subtracted matrix elements.  The results
from different ensembles will then enable an assessment of discretization
and quark-mass tuning effects.  We expect the dominant uncertainty to be due
to the $O(\alpha_s)$ difference between lattice and continuum regularization
schemes.  This will be the first time these matrix elements have been
computed using lattice QCD.

\section*{Acknowledgments}

We thank the MILC collaboration for their gauge configurations and
their code MILC-7.7.11 \cite{MILCgithub}.  This work was funded in
part by STFC grants ST/L000385/1 and ST/L000466/1.  CJM is supported
in part by the US Department of Energy through Grant Number
DE-FG02-00ER41132.  Results described here were obtained using the
Darwin Supercomputer of the University of Cambridge High Performance
Computing Service as part of the DiRAC facility jointly funded by
STFC, the Large Facilities Capital Fund of BIS and the Universities of
Cambridge and Glasgow. This equipment was funded by BIS National
E-infrastructure capital grant (ST/K001590/1), STFC capital grants
ST/H008861/1 and ST/H00887X/1, and STFC DiRAC Operations grants
ST/K00333X/1, ST/M007073/1, and ST/P002315/1.  MW is grateful for an
IPPP Associateship held while this work was undertaken and for
discussions with A Lenz.


\clearpage
\bibliography{mbw}

\end{document}